\documentclass[a4paper,11pt]{article}
\usepackage{latexsym}
\usepackage {amsfonts}
\usepackage[english]{babel}
\font\fr=eufm10 scaled \magstep 1 

\newtheorem{prop}{Proposition}
\newtheorem{corol}{Corollary}

\newtheorem{definition}{Definition}

\def\beq{\begin{equation}}
\def\eeq{\end{equation}}
\def\bea{\begin{eqnarray}}
\def\eea{\end{eqnarray}}
\def\beann{\begin{eqnarray*}}
\def\eeann{\end{eqnarray*}}
\def\beasn{\begin{sneqnarray}}
\def\eeasn{\end{sneqnarray}}
\def\ben{\begin{enumerate}}
\def\een{\end{enumerate}}
\def\bit{\begin{itemize}}
\def\eit{\end{itemize}}
\def\dst{\(\displaystyle}
\def\proof{ {\textbf Proof:} }
\def\derpar#1#2{\frac{\partial{#1}}{\partial{#2}}}

\def\qed{\ifvmode\removelastskip\fi
{\unskip\nobreak\hfil\penalty50\hbox{}\nobreak\hfil
\hbox{\vrule height1.2ex width1.2ex}\parfillskip=0pt
\finalhyphendemerits=0 \par\smallskip}}

\def\vf{\mbox{\fr X}}
\def\df{{\mit\Omega}}
\def\Lag{{\mathfrak L}}

\def\d{{\rm d}}

\def\real{\mbox{\es R}}
\def\real{\mathbb{R}}

\def\Tan{{\rm T}}

\def\inn{\mathop{i}\nolimits}
\def\Cinfty{{\rm C}^\infty}

\renewcommand{\neq}{=\hspace{-3.5mm}/\hspace{2mm}}

\def\tabaddress#1{{\small\it\begin{tabular}[t]{c}#1 \\[1.2ex]\end{tabular}}}

\def\UPCMAT{\it Departamento de Matem\'atica Aplicada IV\\
   Edificio C-3, Campus Norte UPC\\
   C/ Jordi Girona 1. E-08034 BARCELONA,  SPAIN}

\title{TIME-DEPENDENT LAGRANGIANS INVARIANT BY A VECTOR FIELD }
\author{\sc Miguel C. Mu\~noz-Lecanda
\thanks{{\bf e}-{\it mail}: MATMCML@MAT.UPC.ES},
   \\
\sc Narciso Rom\'an-Roy
\thanks{{\bf e}-{\it mail}: MATNRR@MAT.UPC.ES},
    \\
\sc F. Javier Y\'aniz-Fern\'andez 
\thanks{{\bf e}-{\it mail}: YANIZ@MAT.UPC.ES}
   \\
   \tabaddress{\UPCMAT}}
\date{{\sl Letters in Mathematical Physics} {\bf 57} (2001) 107-121}

\begin{document}

\maketitle

\begin{abstract}
We study the reduction of non-autonomous regular Lagrangian systems
by symmetries, which are generated by vector fields associated with connections in the
configuration bundle of the system $Q\times\real\to\real$. These kind of symmetries
generalize the idea of ``time-invariance''
(which corresponds to taking the trivial connection in the
above trivial bundle). 
\end{abstract}


\bigskip
\bigskip
\vbox{\raggedleft AMS s.\,c.\,(2000):
34C14, 37C10, 37J15, 53C05, 70H03
\\
PACS (1999): 02.40.Vh, 45.10.Na, 45.20.Jj}\null

\section{Introduction}

When dealing with mechanical systems coming from a regular  Lagrangian, different formulations exist: If the Lagrangian is autonomous, then the symplectic formalism is the appropriate one, but if the Lagrangian depends on time, then the contact formulation is more suitable. Contact structures are built on  odd dimensional manifolds, and symplectic ones on even dimensional manifolds, which are the phase space of coordinate-velocities.

But these are not the only differences between both formulations. In the symplectic case, the geometric structure is given by a non-degenerate differential 2-form, whereas in the case of the contact, the structural 2-form is degenerate and its kernel has dimension one. This kernel gives at every point the direction of the dynamical vector field.

Obviously, one may use the contact formulation for an autonomous Lagrangian. Then the symmetry given by this time independence allows us to reduce the system, making quotient by the action of time, reducing the dimension of the manifold where the dynamical equations are written, and obtaining the symplectic formulation .

In this work we study the reduction of the contact formulation for a time dependent Lagrangian if there exists an infinitesimal symmetry of the Lagrangian.

The ideas come from  \cite{EMN1,MS,MSA,MW}. There is a  description of the changes in the classical formulation of Lagrangian systems when the natural connection on the bundle
$\pi :Q \times \real \to \real$, given by the vector field $\partial /\partial t$, is replaced by any other  connection,  given by a vector field $\partial /\partial t + Y$, where $Y$ is a vector field on  $Q$ along the projection $\pi _2 :Q \times \real \to Q$. The natural notion of energy associated to any connection of that kind and  its variation along the integral curves of the dynamical vector field of the system is also studied, in particular its invariance, if the Lagrangian is invariant by the action of the vector field associated with the connection. 

In this last situation,  where the vector field associated with the connection is a symmetry of the Lagrangian,  we can mimic the full process of the autonomous case in the contact formulation and obtain a symplectic formulation in an appropriate manifold, given by the quotient of $\Tan Q \times \real$ by the action of the vector field of the connection.  So, every time we have a time dependent regular Lagrangian with a suitable symmetry, the system can be reduced and the symplectic formulation used.

The organization of the paper is as follows: In paragraph 2, we summarize the geometric foundations used in the sequel for non autonomus Lagrangian regular systems. We introduce a non standard connection in the configuration bundle and consider the consequences of this choice. Paragraph 3 is devoted to a study of the relation between symmetries and connections. In paragraph 4 we study the reduction of the system, in particular the projectability of the geometrical structures in order to construct a dynamical system on the quotient manifold.
Paragraph 5 is devoted to studying the properties of the reduced system, which is  a Hamiltonian  but not a Lagrangian system, although the phase space is canonically diffeomorphic to a tangent bundle. The paper ends with some examples and conclusions.

Throughout the paper, every manifold, function and mapping is assumed to be smooth. Manifolds are assumed to be connected and second countable. Summation over crossed repeated indices is understood.

\section{Time-dependent Lagrangian systems}

\subsection{The 1-jet bundle of $\pi\colon Q\times\real\to\real$.
Geometric structures and connections}

The ideas in this section are known. We merely emphasize the differences
between the general situation and this particular one
in order to make the paper more readable and selfcontained.
See \cite{GHV-72} and \cite{S} as general references.

Consider the bundle $\pi\colon Q\times\real\to\real$,
where $Q$ is an $n$-dimensional differentiable manifold
(the {\sl configuration space} of a physical system).
The $1$-jet bundle of sections of $\pi$ is
$\pi_1\colon \Tan Q\times\real\to Q\times\real$.

\begin{prop}
The following elements on  $\pi\colon Q\times\real\to\real$
can be canonically constructed one from the other:
\ben
\item
A section of $\pi_1\colon\Tan Q\times\real\to Q\times\real$,
that is a mapping $\nabla\colon Q\times\real\to\Tan Q\times\real$
such that $\pi_1\circ\nabla ={\rm Id}_{Q\times\real}$.
\item
A subbundle ${\rm H}(\nabla )$ of $\Tan (Q\times\real )$
such that
\beq
\Tan (Q\times\real )={\rm V}(\pi )\oplus{\rm H}(\nabla )
\label{split}
\eeq
\item
A semibasic $1$-form $\tilde\nabla$ on $Q\times\real$
with values in $\Tan (Q\times\real )$ (that is, an element of
${\mit {\mit \Gamma}} (Q\times\real ,\pi^*\Tan^*Q)\otimes\vf (Q\times\real )$),
such that $\alpha\circ\tilde\nabla =\alpha$, for every
semibasic form $\alpha\in\df^1(Q\times\real )$.
\een
\end{prop}
(We denote by $\Gamma (A,B)$ for the set of sections
of the bundle $B\to A$).

\begin{definition}
A {\rm connection} in the bundle $\pi\colon Q\times\real\to\real$
is one of the above mentioned equivalent elements. $\nabla$ is usually called a jet field.
 ${\rm H}(\nabla )$ is called the {\rm horizontal subbundle} of
$\Tan (Q\times\real )$ associated with $\nabla$
and its sections {\rm horizontal vector fields}.
Finally $\tilde\nabla$ is called the Ehresmann connection form.
\end{definition}

Given the subbundle ${\rm H}(\nabla )$ and
the splitting (\ref{split}), we have the maps
$$
h_{\nabla}\colon\Tan (Q\times\real )\to {\rm H}(\nabla )
\ , \ 
v_{\nabla}\colon\Tan (Q\times\real )\to {\rm V}(\pi )
$$
called the {\sl horizontal} and {\sl vertical projections}
(we will use the same symbols $h_{\nabla}$ and $v_{\nabla}$
for the natural extensions of these maps to vector fields).

In a local chart $(q^\mu ,t,v^\mu )$ the expressions
of all these elements are
$$
\nabla (q,s)=(q,s,{\mit \Gamma}^\mu(q,s))
\quad , \quad
\tilde\nabla =
\d t\otimes\left(\derpar{}{t}+{\mit \Gamma}^\mu\derpar{}{q^\mu}\right)
\quad , \quad
{\rm H}(\nabla ) =
{\rm span}\left\{\derpar{}{t}+{\mit \Gamma}^\mu\derpar{}{q^\mu}\right\}
$$
(for every $(q,s)\in Q\times\real$). 

\begin{prop}
A connection in the bundle $\pi\colon Q\times\real\to\real$
is equivalent to a time-dependent vector field in $Q$, that is a vector field along $\tau_{Q}: Q \times \real \longrightarrow Q $.
\label{vfc}
\end{prop}

Given  a connection $\nabla$,let $Y$ be its associated vector field. The {\sl suspension} of $Y$ is the vector field
$ \tilde Y:= \derpar{}{t}+Y $.

In the Lagrangian formalism, the dynamics takes place in the manifold
$\Tan Q\times\real$.
Then, in order to set the dynamics, we need to introduce some geometrical elements
in the bundle $\pi_1: \Tan Q\times\real\to Q\times\real$
(see \cite{EMR-91}, \cite{Gc-74}, \cite{GS-73} and \cite{S}
for details). Every time it is required, we will use a local system given by
$(q^{\mu},t, v^{\mu})$.

We can define a 1-form $\vartheta$ in $\Tan Q\times\real$,
with values in $\pi_1^*{\rm V}(\pi )$, in the following way:
$$
\vartheta (((q,s),u);X):=(\Tan _{((q,s),u)}\pi_1-
\Tan_{((q,s),u)}(\phi\circ\pi\circ\pi_1))(X_{(q,s)})
$$
where $\phi$ is a representative of $((q,s),u)\in\Tan Q\times\real$.
$\vartheta$ is called the {\sl structure canonical form} of
$\Tan Q\times\real$. Its local expression is
\ \dst
\vartheta = (\d q^\mu-v^\mu\d t) \otimes\derpar{}{q^\mu}
\).

Taking into account that $\pi_1\colon\Tan Q\times\real\to Q\times\real$
is a vector bundle, and the fiber on $(q,s)\in Q\times\real$
is $\Tan_qQ\times\{s\}$, there exists a canonical diffeomorphism
between the $\pi_1$-vertical subbundle and $\pi_1^*(\Tan Q\times\real )$,
that is,
$$
{\rm V}(\pi_1)\simeq\pi_1^*(\Tan Q\times\real )
\simeq\pi_1^*\pi_Q^*\Tan Q\simeq\pi_1^*{\rm V}(\pi )
$$
We denote by
$
{\cal S}:\pi_1^*{\rm V}(\pi )\to {\rm V}(\pi_1)
$ \ 
the realization of this isomorphism, and we will use the same notation
${\cal S}$ for its action on the modules of sections of these bundles.
${\cal S}$ is an element of
${\mit \Gamma} (\Tan Q\times\real,\pi_1^*{\rm V}^*(\pi ))
\otimes{\mit \Gamma} (\Tan Q\times\real,{\rm V}(\pi_1))$.
Taking into account that the structure form $\vartheta$ is an element of
${\mit\Omega}^1(\Tan Q\times\real ,\pi_1^*{\rm V}(\pi_1)) =
{\mit\Omega}^1(\Tan Q\times\real )\otimes
{\mit \Gamma} (\Tan Q\times\real,\pi_1^*{\rm V}(\pi ))$,
using the natural duality, by contracting ${\cal S}$ with $\vartheta$,
we obtain an element
$$
{\cal V} := \inn ({\cal S})\vartheta \in
{\mit\Omega}^1(\Tan Q\times\real ) \otimes
{\mit \Gamma} (\Tan Q\times\real,{\rm V}(\pi_1 ))
$$
whose local expression is
\dst
{\cal V}=\left(\d q^\mu-v^\mu\d t\right)\otimes\derpar{}{v^\mu}
\).
Notice that ${\cal V}$ can be thought of as a
$\Cinfty (\Tan Q\times\real )$-module
morphism ${\cal V}\colon\vf (\Tan Q\times\real)\to\vf (\Tan Q\times\real)$
with image on the $\pi_1$-vertical vector fields.

${\cal S}$ and ${\cal V}$ are called the
{\sl vertical endomorphisms} of $\Tan Q\times\real$.
  
\subsection{Lagrangian formalism. Connections and Lagrangian energy functions}

A {\sl time-dependent Lagrangian function} is a function
$L \in C^{\infty} (\Tan Q\times\real )$.

\begin{definition}
The {\rm Poincar\'e-Cartan $1$ and $2$-forms}
associated with the Lagrangian function $L$
are the forms in $\Tan Q\times\real $ defined by
$$
\Theta_{L} := \d L\circ{\cal V}+L \d t
\quad , \quad
\Omega_{L} := -\d\Theta_{L}
$$
\end{definition}

The coordinate expressions of the Poincar\'e-Cartan forms are
\beann
\Theta_{L} &=&
\derpar{L}{v^\mu}(\d q^\mu-v^\mu\d t)+{L}\d t =
\left(L -v^\mu\derpar{L}{v^\mu}\right)\d t+
\derpar{L}{v^\mu}\d q^\mu
\\
\Omega_{L}&=&
-\d\left(\derpar{L}{v^\mu}\right)\wedge\d q^\mu +
\d\left(\derpar{L}{v^\mu}v^\mu -L \right)\wedge\d t
\eeann
Observe that these elements do not depend on the connection.

A Lagrangian function
$L$ is {\sl regular \/} if its associated form
$\Omega_{L}$ has maximal rank, which
is equivalent to demanding that
\dst{\rm det}
\left(\frac{\partial^2 L}{\partial{v^\mu}\partial{v^\nu}}\right)\)
is different from zero at every point.

On the other hand, if we have a connection $\nabla$ on $\pi: Q \times \real \longrightarrow \real$, we can split $\Theta_{L}$ and $\Omega_{L}$ into a sum of vertical and horizontal forms. The vertical form (resp. horizontal) is characterized by the fact that it vanishes under the action of the horizontal (resp. vertical) vector fields associated with the connection. 

Thus $\Omega_{L}= \Omega_{L}^{H} + \Omega_{L}^{V}$ where $\Omega_{L}^{H}=\d t \wedge i(j^{1} \widetilde{Y}) \Omega_{L}$ and $\Omega_{L}^{V}= \Omega_{L} - \Omega_{L}^{H}$. The spliting for $\Theta_{L}$ follows in the same way as above, $\Theta_{L}^{H}= \d t \wedge i(j^{1} \widetilde{Y}) \Theta_{L}$, and $\Theta_{L}^{V}=\Theta_{L} - \Theta_{L}^{H}$.

It is easy to see that this construction can be generalized to every $\alpha \in \Lambda^{r}(F)$ where $F$ is fibred manifold on the basis $B$; see \cite{LMMMR} for more details. 

Assuming the regularity of $L$,
the dynamics of the system is described by a vector field
$X_{L}\in{\cal X}(\Tan Q\times\real )$,
which is a {\sl Second Order Differential Equation} (SODE),
such that:
\beq
\inn(X_{L})\Omega_{L} = 0
\quad , \quad
\inn(X_{L})\d t = 1
\label{dineq}
\eeq
As a consequence, the integral curves of $X_{L}$ verify the
{\it Euler-Lagrange equations}.

A very different picture arises when we try to define intrinsically the
{\sl Lagrangian energy function}. 

\begin{definition}
Let $\nabla$ be a connection in $\pi\colon Q\times\real\to \real$,
$\tilde Y\in\vf (Q\times\real )$ the suspension of the vector field associated with it, and
$j^1\tilde Y\in\vf (\Tan Q\times\real )$ its canonical lifting.
The {\rm Lagrangian energy function} associated with the Lagrangian function
$L$ and the connection $\nabla$ is ${\rm E}^{\nabla}_{L}=-\inn (j^1\tilde Y)\Theta_{L}$
\label{energ}
\end{definition}

In a local chart,
if \dst\tilde Y=\derpar{}{t}+{\mit \Gamma}^\mu\derpar{}{q^\mu}\) , we have
\ \dst
j^1\tilde Y=\derpar{}{t}+{{\mit \Gamma}}^\mu\derpar{}{q^\mu}+
\left(\derpar{{\mit {\mit \Gamma}}^\mu}{t}+v^\nu\derpar{{\mit {\mit \Gamma}}^\mu}{q^\nu}\right)
\derpar{}{v^\mu}
\)\ 
and ${\rm E}_{L}^\nabla = \derpar{L}{v^\mu} (v^\mu-{\mit \Gamma}^\mu )- L$

It is obvious from this expression that
the Lagrangian energy is connection-depending.
If ${\mit \Gamma}^\mu =0$, the corresponding connection is the standard one. The use of this connection ``hides''
the explicit dependence on the connection of the energy
in classical mechanics. The proof of the following propositions can be found in \cite{EMN1}.

\begin{prop}
Let $X_{L}\in\vf (\Tan Q\times\real )$ be the dynamical vector field
(solution of the equations (\ref{dineq})). Then
$$
X_{L}({\rm E}^{\nabla}_{L})=-(j^1\tilde Y) L
$$
\end{prop}

\begin{prop}
If $L$ is a Lagrangian function such that its associated
Legendre transformation
is different from zero at every point,
then every first integral of the dynamical vector field $X_{L}$
is the energy associated with some connection.
\end{prop}

\section{Symmetries of Lagrangian time-dependent systems. Infinitesimal symmetries and connections} 

Let $L \in C^{\infty}(\Tan Q \times \mathbb{R})$ be a time-dependent regular Lagrangian, and $\Phi$ a bundle diffeomorphism  of $Q \times \real$ such that its restriction $\Phi_{\real} : \real \longrightarrow \real$  verifies $\Phi_{\real}(t)=t+c$. $\Phi$ is a {\sl symmetry} if $(j^{1}\Phi)^{\ast}L=L$, that is $L$ is invariant by $j^1 \Phi$.
Let us recall that in this situation the canonical forms $\Theta_{L}$ and $\Omega_{L}$ are also invariant by $j^{1} \Phi$, see \cite{AM}. 

\begin{prop} 
If $L$ is invariant by $j^{1}\Phi$, and $X_L$ is a solution of the equations (\ref{dineq}), then  $(j^{1}\Phi)^{\ast} X_{L}$ is also a solution. 
\end{prop}
\proof
Since $(j^{1}\Phi)^{\ast}L=L$ and $({j^{1}\Phi})^{\ast} \d t= \d( (j^{1}\Phi)^{\ast}\, t)=\d (t+c)= \d t$, we have that

$$0= j^{1}\Phi^{\ast}(i(X_{L}) \Omega_{L})= i(j^{1} \Phi^{\ast}(X_{L}))j^{1}\Phi^{\ast}(\Omega_{L})= i(j^{1}\Phi^{\ast}(X_{L})) \Omega_{L}$$

\noindent and in a similar way
\vspace{-0.3cm}
$$1= j^{1}\Phi^{\ast}(i(X_{L}) \d t)= i(j^{1}\Phi^{\ast}(X_{L})) \d t.$$
\qed

{\bf Remark}: As $L$ is regular, we have that $(j^1 \Phi)^{\ast} X_L= X_L$, hence if $\sigma$ is an integral curve of $X_{L}$, then $\Phi^{\ast}\sigma$ is too.

Given $Y \in \mathfrak{X}(Q, \tau_{Q})$ a vector field along the projection $\tau_{Q}$, consider the vector field $j^1 \widetilde{Y}\in \mathfrak{X}(\Tan Q \times \real)$ where $\widetilde{Y}=\derpar{}{t} + Y$. We say that $j^1 \widetilde{Y}$ is an {\sl infinitesimal symmetry} of the system if $\mathcal{L}(j^1 \widetilde{Y})L=0$, where $\mathcal{L}(X)$ is the Lie derivative operator with respect to $X$.

As a consequence of the above proposition, we have:

\begin{corol} 
If $\mathcal{L}(j^1 \widetilde{Y})L=0$, then $\mathcal{L}(j^1 \widetilde{Y})\Omega_{L}=0$, $\mathcal{L}(j^1 \widetilde{Y})\d t=0$, $\mathcal{L}(j^1 \widetilde{Y})X_L=0$, and $\mathcal{L}(j^1 \widetilde{Y})E^{\nabla}_{L}=0$ .
\end{corol}

\qed

As a particular case, we can consider the invariance under a connection. Let $\nabla$ be a connection with $j^{1} \widetilde{Y} \in \mathfrak{X}(\Tan Q \times \real)$ the 1-jet prolongation of the field $\widetilde{Y}$ associated with $\nabla$, which we will suppose to be complete. The Lagrangian system is called {\sl invariant under the connection} $\nabla$ if $\mathcal{L}(j^{1}\widetilde{Y})L=0$. 

With these conditions, and taking into account that $j^{1}\widetilde{Y}$ is complete, we can define the next family of diffeomorphisms:
\vspace{-0.5cm}
\beann
\widetilde{\Phi}_{s}: \Tan Q \times \real & \longrightarrow & \Tan Q \times \real \\
(v_q,t) & \longrightarrow & j^{1} \varphi_{s}(v_q,t)
\eeann

\noindent where $j^{1} \varphi_{s}$ is the uniparametric flow of $j^{1} \widetilde{Y}$. Observe that if $\mathcal{L}(j^{1} \widetilde{Y})L=0$, then $\widetilde{\Phi}_{s}$ is a symmetry of $L$ for every $s \in \real$. Therefore: 

\begin{prop}
If $\mathcal{L}(j^{1} \widetilde{Y})L=0$ then $\d \Theta_{L}^{V}=- \Omega_{L}^{V}$.
\end{prop}

\proof 
\begin{eqnarray*}
\d \Theta_{L}^{V}&=& \d(\Theta_{L}- i(j^{1} \widetilde{Y}) \Theta_{L} \d t)=- \Omega_{L} - \d ( i(j^{1} \widetilde{Y}) \Theta_{L}) \wedge \d t \\
 &=& -\Omega_{L}- i(j^{1} \widetilde{Y}) \Omega_{L} \wedge \d t + \mathcal{L}(j^{1} \widetilde{Y}) \Omega_{L}= -\Omega_{L} + \d t \wedge i(j^{1} \widetilde{Y}) \Omega_{L}= -\Omega_{L}^{V}
\end{eqnarray*}
\qed

\noindent {\bf Assumption:} From now on, we will suppose that $j^{1}\widetilde{Y}$ is a infinitesimal symmetry of the system $(\Tan Q \times \real, L)$.

\section{Reduction of the time-dependent Lagrangian systems invariant under $\nabla$}

Consider the action of $j^{1} \widetilde{Y}$ on $\Tan Q \times \mathbb{R}$, and let $[\Tan Q \times \real ] \equiv (\Tan Q \times \real) / j^{1} \widetilde{Y}$ be the set of equivalence classes by this action. Let $\overline{\pi}^{1}: \Tan Q \times \real  \longrightarrow [\Tan Q \times \real ]$ the natural projection. We will assume that $[\Tan Q \times \real ]$ is a manifold and $\overline{\pi}^{1}$ a submersion.

Remember that given a projection of manifolds $\tilde{\pi}: M \longrightarrow M / \sim $, and $\omega \in \Lambda^{h}(M)$, we say that $\omega$ is $\widetilde{\pi}-$projectable if there exists $\tilde{\omega}$ such that $\tilde{\pi}^{\ast}(\tilde{\omega})= \omega$. A necessary and sufficient condition to assure the projectability is that $i(X) \omega=0$ and $\mathcal{L}(X) \omega =0$, for every $X \in \mathfrak{X}^{V}(\tilde{\pi})$ or equivalently $i(X)\omega=0, \, i(X) \d \omega=0$, for $X \in \mathfrak{X}^{V}(\tilde{\pi})$ (see \cite{L} for more details). Observe that in our situation $\mathfrak{X}^{V}(\overline{\pi}^{1})$ is spanned by $j^{1} \widetilde{Y}$.

Let $\Omega_{L}^{H}= \d t \wedge i(j^{1}\widetilde{Y})\Omega_{L}$, $\Omega_{L}^{V}= \Omega_{L}- \Omega_{L}^{H}$, and $\Theta_{L}^{H}= (i(j^{1}\widetilde{Y})\Theta_{L}) \, \d t$, $\Theta_{L}^{V}= \Theta_{L}- \Theta_{L}^{H}$ be the decompositions induced by $\nabla$, then we have:

\begin{prop} $\Theta^{V}_{L}$, $\Omega_{L}^{V}$ and $E_{L}^{\nabla}$ are $\overline{\pi}^{1}-$projectable. \end{prop}

\proof  Taking into account that $i(j^{1}\widetilde{Y})\d t=1$, and $i(j^{1}\widetilde{Y})i(j^{1}\widetilde{Y})\Omega_{L}=0$, because $\Omega_{L}$ is skew-symmetric, we have:
\begin{eqnarray*}
i(j^{1}\widetilde{Y}) \Omega^{V}_{L} &=& i(j^{1}\widetilde{Y}) \Omega_{L}- (i(j^{1}\widetilde{Y})\d t)i(j^{1}\widetilde{Y})\Omega_{L}+ (i(j^{1}\widetilde{Y})i(j^{1}\widetilde{Y})\Omega_{L})\d t=0\\
i(j^{1}\widetilde{Y})\d \Omega_{L}^{V}&=& i(j^{1}\widetilde{Y})\d \Omega_{L}- i(j^{1}\widetilde{Y})(\d( \d t) \wedge i(j^{1}\widetilde{Y}) \Omega_{L})+ i(j^{1}\widetilde{Y})(\d t \wedge \d (i(j^{1}\widetilde{Y})\Omega_{L}))\\
                                       &=& i(j^{1}\widetilde{Y})(\d t \wedge \d(i(j^{1}\widetilde{Y})\Omega_{L})) 
\end{eqnarray*}
\noindent where the last expression is equal to zero since $d(i(j^{1}\widetilde{Y})\Omega_{L})= \mathcal{L}(j^{1}\widetilde{Y}) \Omega_{L} - i(j^{1}\widetilde{Y})d\Omega_{L}$. In the same way we can prove the assertion for $\Theta_{L}^{V}$.

For the Lagrangian energy function holds:
\vspace{-0.2cm}
\beann
\mathcal{L}(j^{1}\widetilde{Y}) E_{L}^{\nabla}&=& i(j^{1}\widetilde{Y})E_{L}^{\nabla}= (dE_{L}^{\nabla})(j^{1}\tilde{Y})= -d(i(j^{1}\widetilde{Y}) \Theta_{L})(j^{1}\widetilde{Y}) \\
                      &=&-(i(j^{1}\widetilde{Y})\Omega_{L})(j^{1}\widetilde{Y})=-i(j^{1}\widetilde{Y})i(j^{1}\widetilde{Y})\Omega_{L}=0.
\eeann

\qed

{\bf Remark:} As a consequence, there exist diferential forms $\mathcal{\omega}$ and $\mathcal{\theta}$ in $[ \Tan Q \times \real]$, such that $\overline{\pi}^{ 1 \,\ast}(\mathcal{\omega})= \Omega_{L}^{V}$ and $\overline{\pi}^{1 \,\ast}(\mathcal{\theta})= \Theta_{L}^{V}$, and a function $\mathcal{E} \in C^{\infty}([\Tan Q \times \real])$ verifing that $\overline{\pi}^{1 \,\ast}(\mathcal{E})= E_{L}^{\nabla}$. 
\hspace{0.5cm}

\begin{prop}
$([ \Tan Q \times \real], \mathcal{\omega})$ is a symplectic manifold and  $([ \Tan Q \times \real], \mathcal{\omega}, \mathcal{E})$ is a Hamiltonian system.
\end{prop}

\proof We must show that $\mathcal{\omega}$ is closed and non-degenerated.
\begin{itemize}
\item $\overline{\pi}^{1 \, \ast}(\d \mathcal{\omega})= \d (\overline{\pi}^{1 \, \ast} \mathcal{\omega})= \d \Omega_{L}^{V}= \d \Omega_{L} - \d \Omega_{L}^{H} = - \d(\d t \wedge i(j^{1} \widetilde{Y})\Omega_{L})= \d t \wedge \d ( i(j^{1} \widetilde{Y})\Omega_{L}))=0.$

Since $\overline{\pi}^{1}$ is a submersion, the fact that $\overline{\pi}^{1 \, \ast}(\d \mathcal{\omega})= 0$, implies that $\d \mathcal{\omega}=0$.
\item Suppose that $i(Z) \mathcal{\omega}=0$, then  if $X \in\overline{\pi}^{1 \, \ast}(Z)$, then $0=i(X) \Omega_{L}^{V}=i( X^{V}+ X^{H}) \Omega_{L}^{V}= i(X^{V}) \Omega_{L}^{V}= i(X^{V}) \Omega_{L}$.

Since, $L$ is regular, then $X^{V}=0$, which implies that $Z=0$. 

Observe that $\overline{\pi}^{1 \, \ast}(Z)$ is well-defined because $\overline{\pi}^{1}$ is an exhaustive submersion.

\end{itemize} 
\qed

Considering the dynamical equations in $\Tan Q \times \real$, we write them in a suitable way to obtain the dynamics in the quotient.
\beann
0 &=& i(X_{L}) \Omega_{L}= i(X_{L})(\Omega_{L}^{H} + \Omega_{L}^{V})= i(j^{1} \widetilde{Y}) \Omega_{L} - \d t ( i(X_{L}) i(j^{1} \widetilde{Y}) \Omega_{L}) + i(X_{L}) \Omega_{L}^{V} \\
  &=& i(j^{1} \widetilde{Y}) \Omega_{L} + i(X_{L}) \Omega_{L}^{V}= -dE_{L}^{\nabla} + i(X_{L}) \Omega_{L}^{V}.
\eeann

\noindent where we have taken into account that  $i(X_{L}) i(j^{1} \widetilde{Y}) \Omega_{L}=0$, since $X_{L}$ is a solution of (\ref{dineq}), and $i(j^{1} \widetilde{Y}) \Omega_{L}=- \mathcal{L}(j^{1} \widetilde{Y}) \Theta_{L} + d(i(j^{1} \widetilde{Y}) \Theta_{L})=-\d E_{L}^{\nabla}$.

As the connection $\widetilde{\nabla}$ associated with $j^{1} \widetilde{Y}$ allows us to split $\Tan (\Tan Q \times \real)=H(\nabla)\oplus V(\pi \circ \pi^{1})$, then $X_{L}=X_{L}^{H}+X_{L}^{V}$, and
\beann
0&=& -\d E_{L}^{\nabla} + i(X_{L}^{H}+X_{L}^{V}) \Omega_{L}^{V}= -\d E_{L}^{\nabla} + i(X_{L}^{H}) \Omega_{L}^{V} +i(X_{L}^{V}) \Omega_{L}^{V} = -\d E_{L}^{\nabla} + i(X_{L}^{V}) \Omega_{L}^{V}.
\eeann
\noindent since $i(f j^{1} \widetilde{Y}) \Omega_{L}^{V}= f i(j^{1} \widetilde{Y})\Omega^{V}_{L}=0$. Observe that
the second equation does not give us any information about $X_{L}^{V}$, because $i(X_{L}^{V}) \d t=0$ holds. Then we have $i(X_{L}^{V}) \Omega_{L}^{V}= \d E_{L}^{\nabla}$.

As a consequence of the previous propositions
\begin{eqnarray*}
0= - \d ( \overline{\pi}^{1 \,\ast}(\mathcal{E}))+ i(X_{L}^{V})\overline{\pi}^{1 \,\ast} \mathcal{\omega}
= \overline{\pi}^{1 \,\ast}(-\d \mathcal{E})+ \overline{\pi}^{1 \,\ast}(i(\mathcal{X}) \mathcal{\omega})
= \overline{\pi}^{1 \,\ast}( -\d \mathcal{E}+ i(\mathcal{X}) \mathcal{\omega}).
\end{eqnarray*}

\noindent This implies that $0= -\d \mathcal{E}+ i(\mathcal{X}) \mathcal{\omega}$, because $\overline{\pi}^{1}$ is a submersion. So the dynamical equation in $[\Tan Q \times \real]$ is $ i(\mathcal{X}) \mathcal{\omega}=\d \mathcal{E}$. 

\begin{prop} 
$\mathcal{E}$ is a first integral of the system $([ \Tan Q \times \real], \mathcal{\omega}, \mathcal{E})$. 
\end{prop}
\proof Since $(j^{1} \widetilde{Y}) L=0$ the energy function $E_{L}^{\nabla}$ is constant along the trajectories of (\ref{dineq}). Otherwise $X_{L}^{H}(E_{L}^{\nabla})=0$ because $(\d  E_{L}^{\nabla}) X_{L}^{H}= -i(f j^{1} \widetilde{Y}) i(j^{1} \widetilde{Y}) \Omega_{L}=0$. As a consequence of this $0=X_{L}(E_{L}^{\nabla})= X_{L}^{H}(E_{L}^{\nabla})+ X_{L}^{V}(E_{L}^{\nabla})= X_{L}^{V}(E_{L}^{\nabla})$. Then we have that
$0= X_{L}^{V}(E_{L}^{\nabla})= \pi^{\ast}( \mathcal{X}(\mathcal{E}))$ and $\mathcal{X}(\mathcal{E})=0$.

\qed

{\bf Remark:} The restriction of the Hamiltonian system
 $([\Tan Q \times \real], \mathcal{\omega}, \mathcal{E})$
 to a hypersurface defined by $\mathcal{E}=ctn$,
 gives the same result as if we apply the
 presymplectic reduction procedure, studied in \cite{EMR-99},
 to the initial Lagrangian system $(\Tan Q \times \real, L)$
 under the action of $j^{1} \widetilde{Y}$. 
\vspace{0.5cm}
 
Now suppose that $\mathcal{X} \in \mathfrak{X}([\Tan Q \times \real])$ is the solution of the dynamical equation $i(\mathcal{X}) \mathcal{\omega}=\d \mathcal{E}$. 
We want to recover the solution $X_{L} \in \mathfrak{X}(\Tan Q \times \real)$ of (\ref{dineq}) from $\mathcal{X}$. Let $Z:= j^{1} \widetilde{Y} + v_{\nabla}(\overline{\pi}^{1 \, \ast}(\mathcal{X})) \in \mathfrak{X}(\Tan Q \times \real) $. It is well defined because every $X\in \mathfrak{X}(\Tan Q \times \real)$ such that $\overline{\pi}^{1 \, \ast}(\mathcal{X})=X$ has the same vertical part. Then:

\begin{prop}
$Z \in \mathfrak{X}( \Tan Q \times \real)$ is a solution of (\ref{dineq}).
\end{prop}
\proof As $\d t$ is a semibasic form we have:
\begin{eqnarray*}
i(Z)\d t &=& i(j^{1}\widetilde{Y}) \d t + i(v_{\nabla}(\overline{\pi}^{1 \, \ast}(\mathcal{X}))) \d t = 1+ i(v_{\nabla}(\overline{\pi}^{1 \, \ast}(\mathcal{X}))) \d t=1 \\
i(Z) \Omega_{L}&=& i(j^{1} \widetilde{Y}) \Omega_{L}+ i(v_{\nabla}(\overline{\pi}^{1 \, \ast}(\mathcal{X}))) \Omega_{L}^{V}= i(j^{1} \widetilde{Y}) \Omega_{L} + i(\overline{\pi}^{1 \, \ast}(\mathcal{X})) \Omega_{L}^{V} \\                 &=& - \overline{\pi}^{1 \, \ast}(\d \mathcal{E}) + i(\overline{\pi}^{1 \, \ast}(\mathcal{X})) \overline{\pi}^{1 \, \ast}\omega= - \overline{\pi}^{1 \, \ast}(\d \mathcal{E})+ \overline{\pi}^{1 \, \ast}(i(\mathcal{X}) \omega)= \overline{\pi}^{1 \, \ast}( -\d \mathcal{E}+ i(\mathcal{X}) \omega)=0 
\end{eqnarray*}
\qed

\section{Properties of the system $([ \Tan Q \times \real ], \omega, \mathcal{E})$}

Consider now the manifold $[ Q \times \real]:= (Q \times \real)/ \widetilde{Y}$ and the submersion $\overline{\pi}: Q \times \real \longrightarrow [ Q \times \real]$. From every element $[(q,t)] \in [Q \times \real ]$ we can choose a representative of the form $(\tilde{q}, 0)$: if $(q,t)\in [(q,t)]$ then, $(\tilde{q},0)= \varphi_{-t}(q,t)$ where $\varphi_{s}(q,t)=(\varphi^{Q}_{s}(q,t), t+s)$ is the flow of $\widetilde{Y}$. This element is unique, since if there exists $(q^1,0)$ and $(q^2, 0)$ in the same class $[(q,t)]$, then for some $s \in \real$, we have $(q^2,0)=(\varphi^{Q}_{s}(q^1,0),s)$. Hence we can conclude that $s=0$ and thus $q^1=q^2$.

 From the above considerations, there exists a natural bijection $\psi$ from $[Q \times \real]$ to $Q$.
$$
\begin{array}{ccccccccc}
\psi:& [Q \times \real] & \longrightarrow & Q & \quad \quad & \psi^{-1}:& Q & \longrightarrow & [Q \times \real] \\
   & \lbrack (q,t) \rbrack & \longrightarrow & \varphi^{Q}_{-t}(t,q)&\quad \quad &   & q & \longrightarrow & [(q,0)]
\end{array}
$$
It is clear that $\psi \circ \psi^{-1}=Id \vert_{Q}$ and $\psi^{-1} \circ \psi=Id \vert_{[Q \times \real]}$.

\begin{prop}
$\psi$ is a diffeomorphism.
\end{prop}
\proof Consider the diagram

$$\begin{array}{cc}
 Q \times \real 
\quad
\begin{picture}(50,10)(0,0)
\put (0,5) {\vector(1,0){50}}
\put(20,7){ \mbox{$\overline{\pi}$}}
\end{picture}
& [Q \times \real]
\\

\begin{picture}(60,40)(0,0)
\put (5,45) {\vector(2,-1){90}}
\put(30,15){ \mbox{$\phi$}}
\end{picture}
&
\begin{picture}(10,50)(0,0)
\put (5,40) {\vector(0,-1){40}}
\put (10,0) {\vector(0,1){40}}
\put(12,20){ \mbox{$\psi^{-1}$}}
\put(-12,20){ \mbox{$\psi$}}
\end{picture} 
\\
    & Q 
\end{array}$$

\noindent where $\phi(t,q)= \varphi^{Q}_{t}(t,q)$. $\psi$ is smooth iff $\phi$ is smooth, see \cite{GHV-72}, and $\phi$ is smooth trivially.

On the other hand, we can see $Q$ as a quotient manifold of $Q \times \real$, then using the same pattern as above, $\psi^{-1}$ is smooth because $\overline{\pi}$ is too. Therefore $\psi$ is a diffeomorphism.
\qed

Observe that by means of $\psi$ we can build natural local charts on $[Q \times \real]$. 
Let $\{ U_{\alpha}, \zeta_{\alpha} \}$ one atlas on $Q$, then it is easy to see that $\{ \psi^{-1}(U_{\alpha}), \zeta_{\alpha} \circ \psi \}$ is an atlas on $[Q \times \real]$.
\vspace{0.3cm}

{\bf Remark:} As a consequence of the above considerations, we can also construct a diffeomorphism $$\widetilde{\psi}: [\Tan Q \times \real] \longrightarrow \Tan Q$$
Now we can define the projecction map $\overline{\pi}_{1}: [ \Tan Q \times \real] \longrightarrow [ Q \times \real ]$, because given an integral curve, $\gamma$, of $j^{1} \widetilde{Y}$, $\pi_{1}\circ \gamma$ is an integral curve of $\widetilde{Y}$. Then we have:
\begin{eqnarray*}
\overline{\pi}_{1}: [\Tan Q \times \real] & \longrightarrow & [Q \times \real ]\\
 \lbrack (v_q, t) \rbrack & \longrightarrow & [(q,t)]
\end{eqnarray*}

\begin{prop}
$\overline{\pi}_{1}$ is a submersion.
\end{prop}

\proof In natural local charts, $\overline{\pi}_{1}(t,q,v)=(t,q)$, hence $\overline{\pi}_{1}$ is smooth and surjective.

\qed
 From the previous considerations, the following diagram is commutative:
$$\begin{array}{ccccc}
\Tan Q \times \real 
&
\begin{picture}(50,10)(0,0)
\put (0,5) {\vector(1,0){50}}
\put(20,7){ \mbox{$\overline{\pi}^{1}$}}
\end{picture}
&
[ \Tan Q \times \real]
&
\begin{picture}(50,10)(0,0)
\put (0,5) {\vector(1,0){50}}
\put(20,8){ \mbox{$\widetilde{\psi}$}}
\put(20,-8){ \mbox{$\cong$}}
\end{picture}
&
 \Tan Q \\

\begin{picture}(10,50)(0,0)
\put(5,50){\vector(0,-1){50}}
\put (-12,20) { \mbox{$\pi_{1}$}}
\end{picture}
& &
\begin{picture}(10,50)(0,0)
\put(5,50){\vector(0,-1){50}}
\put (-12,20) { \mbox{$\overline{\pi}_{1}$}}
\end{picture}
& &
\begin{picture}(10,50)(0,0)
\put (5,50){\vector(0,-1){50}}
\put (-12,20) { \mbox{$\tau_{Q}$}}
\end{picture}
\\

Q \times \real
&

\begin{picture}(50,10)(0,0)
\put (0,5) {\vector(1,0){50}}
\put(20,7){ \mbox{$\overline{\pi}$}}
\end{picture}
& [ Q \times \real ] &

\begin{picture}(50,10)(0,0)
\put (0,5) {\vector(1,0){50}}
\put(20,8){ \mbox{$\psi$}}
\put(20,-8){ \mbox{$\cong$}}
\end{picture}
& Q \\
\end{array}
$$

Since $L \in C^{\infty}(\Tan Q \times \real)$ is constant along the integral curves of $j^{1} \widetilde{Y}$, we can define a function $\Lag$ on $[\Tan Q \times \real]$. Consider the Lagrangian forms $\theta_{\bar{\Lag}}$ and $\omega_{\bar{\Lag}}$ on $\Tan Q$ associated with the Lagrangian function $\bar{\Lag}= \widetilde{\psi}_{\ast}(\Lag)$. We wish to compare them with $ \widetilde{\psi}_{\ast}\omega$ and $\widetilde{\psi}_{\ast} \theta $.

\begin{prop}
$(\widetilde{\psi} \circ \overline{\pi}^{1})^{\ast} ( \theta_{\bar{\Lag}})= \Theta_{L}^{V} = \overline{\pi}^{1 \, \ast}( \theta)$.
\end{prop}
\proof It is enough to show this proposition in local coordinates. Let $\phi:=\widetilde{\psi} \circ \overline{\pi}^{1}$ be, as $$\phi(t,q^i,v^i)= \left( \varphi_{-t}^{i}(q^i,t),\derpar{\varphi^{i}_{s}}{t}(-t;(q^i,t))+ \derpar{\varphi_{s}^{i}}{q^{j}}(-t;(q^i,t))  \, v^{j} \right)=:(\overline{q}^{i}, \bar{v}^{i})$$
\noindent and $L(t,q^i,v^i)= \bar{\Lag}(\phi(t,q^i,v^i))$, we have that

\begin{eqnarray*}
 (\widetilde{\psi} \circ \overline{\pi}^{1})^{\ast}(\theta_{\bar{\Lag}}) &=& \left. \derpar{\bar{\Lag}}{\bar{v}^{i}}\right \vert_{\phi(t,q^i,v^i)} \left( \left. \derpar{\varphi_{-t}^{i}}{q^{j}}\right \vert_{(t,q^i)}\, \d q^{j} + \left. \derpar{\varphi_{-t}^{i}}{t} \right \vert_{(t,q^i)} \, \d t \right)= 
  \left. \left( \derpar{\bar{\Lag}}{\bar{v}^{i}}\right \vert_{\phi(t,q^i,v^i)} \left.\derpar{\varphi_{-t}^{i}}{q^j} \right \vert_{(t,q^i)} \right) \d q^j + \\
 &=& \left. \derpar{\bar{\Lag}}{\bar{v}^{i}}\right \vert_{(\phi(t,q^i,v^i))} \left. \derpar{\varphi_{-t}^{i}}{t} \right \vert_{(t,q^i)} \d t 
 = \left. \derpar{L}{v^j}\right \vert_{(t,q^i, v^i)} \, \d q^j + \left. \derpar{\bar{\Lag}}{\bar{v}^{i}}\right \vert_{\phi(t,q^i,v^i)} \left. \derpar{\varphi_{-t}^{i}}{t} \right \vert_{(t,q^i)} \, \d t + \\
  & & \left. \derpar{\bar{\Lag}}{\bar{v}^{i}} \right \vert_{\phi(t,q^i,v^i)} \left(  - {\mit {\mit \Gamma}}^{j}(q,t) \left. \derpar{\varphi_{-t}^{i}}{q^j}\right \vert_{(t,q^i)} \,+  {\mit {\mit \Gamma}}^{j}(q^i,t) \left. \derpar{\varphi_{-t}^{i}}{q^j} \right \vert_{(t,q^i)} \right) \d t \\
&=& \left. \derpar{L}{v^j}\right \vert_{(t,q^i, v^i)} \, \d q^j + 
 \left. \derpar{\bar{\Lag}}{\bar{v}^{i}}\right \vert_{\phi(t,q^i,v^i)} \left( - {\mit {\mit \Gamma}}^{j}(q,t) \left. \derpar{\varphi_{-t}^{i}}{q^j}\right \vert_{(t,q^i)} \, \right) \d t + \\
 & & \left. \derpar{\bar{\Lag}}{\bar{v}^{i}} \right \vert_{\phi(t,q^i,v^i)} \left( \left. \derpar{\varphi_{-t}^{i}}{t}\right \vert_{(t,q^i)} \,+ {\mit {\mit \Gamma}}^{j}(q^i,t) \left. \derpar{\varphi_{-t}^{i}}{q^j} \right \vert_{(t,q^i)} \right) \d t  \\
&=& \left. \derpar{L}{v^j}\right \vert_{(t,q^i, v^i)} \d q^j - {\mit {\mit \Gamma}}^{j}(q^i,t) \left. \derpar{L}{v^j}\right \vert_{(t,q^i, v^i)} \d t = \Theta_{L}^{V}
\end{eqnarray*}

\noindent where we have taken into account that 
$$\left. \derpar{L}{v^i} \right \vert_{(t,q^i, v^i)}= \left. \derpar{\bar{\Lag}}{\bar{v}^{j}} \right \vert_{\phi(t,q^i,v^i)} \left. \derpar{\varphi^{j}_{s}} {q^i}\right \vert_{(-t;(t,q^i))}= \left. \derpar{\bar{\Lag}}{\bar{v}^j}\right \vert_{\phi(t,q^i, v^i)} \left. \derpar{\varphi_{-t}}{q^j} \right \vert_{(t,q^i)}.$$

Observe that $\left. \derpar{\varphi_{-t}^{i}}{t} \right \vert_{(t,q^i)}+ \left. {\mit {\mit \Gamma}}^{j}\right \vert_{(t,q^i)} \left. \derpar{\varphi_{-t}^{i}}{q^j}\right \vert_{(t,q^i)}=0$, since the tangent mapping of $(t,q^i)\mapsto (t, \varphi^{i}_{-t}(q^i,t))$ transforms the vector field $\derpar{}{t}+ {\mit {\mit \Gamma}}^{i}(q^i,t) \derpar{}{q^{i}}$ into $\derpar{}{t}$.

\qed
{\bf Remark:} The above result is due to the fact that the symmetry is natural, that is, a jet prolongation of a vector field on $Q \times \real$.

\begin{prop}
$\overline{\pi}^{1 \, \ast}( \widetilde{\psi}^{\ast}(\omega_{\bar{\Lag}}))= \Omega_{L}^{V}$
\end{prop}
\proof Since $\omega_{\bar{\Lag}}=-\d \theta_{\bar{\Lag}}$, we have that $(\psi \circ \overline{\pi}^{1})^{\ast}(\d \theta_{\bar{\Lag}})= d((\psi \circ \overline{\pi}^{1})^{\ast} \theta_{\bar{\Lag}})=\d \Theta_{L}^{V}$. On the other hand, it verifies that $\d \Theta_{L}^{V}
 = -\Omega_{L}^{V}$.Therefore $\overline{\pi}^{1 \, \ast}( \widetilde{\psi}^{\ast}(\omega_{\bar{\Lag}}))= \Omega_{L}^{V}$.
\qed

\begin{corol}
$\theta= \widetilde{\psi}^{\ast}(\theta_{\bar{\Lag}})$ and $\omega= \widetilde{\psi}^{\ast}( \omega_{\bar{\Lag}})$.
\end{corol}
\proof
>From the above proposition we have that $\overline{\pi}^{1 \, \ast}( \widetilde{\psi}^{\ast}(\theta_{\bar{\Lag}}))= \overline{\pi}^{1 \, \ast}( \theta)$, then, since $\overline{\pi}^{1}$ is a submersion, $\theta= \widetilde{\psi}^{\ast}(\theta_{\bar{\Lag}})$. The second assertion follows in the same way.
\qed
A different situation arises when we try to do the same with the energy. Let $\mathcal{E}_{\bar{\Lag}}:= \Delta \bar{\Lag} - \bar{\Lag}$ be the energy associated with the Lagrangian $\bar{\Lag}$, then in general
$\overline{\pi}^{1 \, \ast}( \widetilde{\psi}^{\ast}(\mathcal{E}_{\bar{\Lag}}))\neq E_{L}^{\nabla}$. An example of this is given in the next section.

\noindent {\bf Remark:} So the Hamiltonian system $([\Tan Q \times \real ], \omega, \mathcal{E})$ is not, in general, a Lagrangian one.

\section{Examples}
\subsection{Autonomous dynamical systems}

First we analyze the time-independent dynamical systems as a particular case of non-autonomous regular systems which are invariant under time translations.

Let $(\Tan Q \times \real, \Omega_{L})$ the non-autonomous regular Lagrangian dynamical system. $\nabla_{0}$ is the standard connection, and as a consequence $\widetilde{Y}= \frac{\partial}{\partial t}$. Let $L \in C^{\infty}(\mathbb{R} \times TQ)$ be the Lagrangian function, such that $\mathcal{L}(j^{1} \widetilde{Y}) L= \mathcal{L}(\frac{\partial}{\partial t}) L=0$, that is $\frac{\partial L}{\partial t}=0$. Then $\varphi(t,q,v_q)=(q,v_q)=:(\bar{q},\bar{v}_{\bar{q}})$, and $\bar{\Lag}(\bar{q}, \bar{v}_{\bar{q}})= L(0,q,v_q)$. Therefore  $\theta =\theta_{\bar{\Lag}}$, $\omega =\omega_{\bar{\Lag}}$ and $\mathcal{E}= \mathcal{E}_{\bar{\Lag}}$. 

Observe that in this particular case the system $([\Tan Q \times \real ], \omega, \mathcal{E})$ is a Lagrangian one.

\subsection{Another example}

Let $(\Tan \real^{2} \times \real, \Omega_{L})$ be the regular lagrangian dynamical system associated with the Lagrangian function $$L(t,x,y,v_x,v_y)= \frac{1}{2} (v_{x}^{2}+v_{y}^{2})- V(y)+ (t-x) v_{x}+ (t-x) v_{y}.$$

Consider the infinitesimal symmetry $j^{1} \widetilde{Y}= \derpar{}{t}+ \derpar{}{x}$. Taking into account the coordinate expression of the flow, we have that $\bar{\Lag}(\bar{x},\bar{y}, \bar{v}_{\bar{x}}, \bar{v}_{\bar{y}})= \frac{1}{2}( \bar{v}_{\bar{x}}^{2}+ \bar{v}^{2}_{\bar{y}})- V(\bar{y})- \bar{x} \, \bar{v}_{\bar{x}} - \bar{x}\, \bar{v}_{\bar{y}}$. From the above considerations $\theta_{\bar{\Lag}}= (\bar{v}_{\bar{x}}- \bar{x}) \, \d \bar{x} + ( \bar{v}_{\bar{y}}- \bar{x}) \, \d \bar{y}$, and as we know
\vspace{-0.3cm}
\begin{eqnarray*}
\varphi^{\ast}(\theta_{\bar{\Lag}})&=&(v_x - x+t) (\d x- \d t)+ (v_y -x + t) \d y= \\
                                        &=& (x-v_x +t) \d t + (v_x -x +t) \d x + (v_y -x +t) \d y= \Theta_{L}^{V}
\end{eqnarray*}

On the other hand, since $E_{L}^{\nabla}= \frac{1}{2}(v_{x}^{2}+ v_{y}^{2})+ V(y) +x -t-v_x$, then $\mathcal{E}= \frac{1}{2}(\bar{v}^{2}_{\bar{x}} + \bar{v}^{2}_{\bar{y}})+ V(\bar{y})+ (\bar{x}- \bar{v}_{\bar{x}})$, which is different from $\mathcal{E}_{\bar{\Lag}}= \frac{1}{2}(\bar{v}^{2}_{\bar{x}} + \bar{v}^{2}_{\bar{y}})+ V(\bar{y})$.

Now we can solve the dynamical equation $i(\mathcal{X}_{\bar{\Lag}}) \omega_{\bar{\Lag}}= \d \mathcal{E}$, obtaining that $$\mathcal{X}_{\bar{\Lag}}= (\bar{v}_{\bar{x}}-1) \derpar{}{\bar{x}}+ \bar{v}_{\bar{y}} \derpar{}{\bar{y}}-(1+ \bar{v}_{\bar{y}}) \derpar{}{\bar{v}_{\bar{x}}}
+ (\bar{v}_{\bar{x}}-1- \derpar{\overline{V}}{\bar{y}} ) \derpar{}{\bar{v}_{\bar{y}}}$$
\noindent and therefore $X_{L}= \derpar{}{t}+v_x \derpar{}{x}+ v_y \derpar{}{y} -(1+v_y) \derpar{}{v_x}+ (v_x -1- \derpar{V}{y}) \derpar{}{v_y}$.

\section{Conclusions and outlook}

Let $(\Tan Q\times\real,L)$ be a time-dependent regular Lagrangian system.
Every connection $\nabla$ in $Q\times\real\to\real$ is associated with a
vector field $\tilde Y\in\vf(Q\times\real)$. Then, we study the
reduction of the system when $j^1\tilde Y\in\vf(\Tan Q\times\real)$
is an infinitessimal symmetry of $L$. First, the connection $\nabla$
allows us to split the Lagrangian 2-form $\Omega_L$
into the corresponding vertical and horizontal parts,
$\Omega_L^V$ and $\Omega_L^H$, and then,
introducing the Lagrangian energy function $E_L^\nabla$
associated with $\nabla$, and using $\Omega_L^V$, we can write the dynamical
equations in a similar way to the time-independent case.
Therefore, considering the reduced manifold
$[\Tan Q\times\real]\equiv\Tan Q\times\real/j^1\tilde Y$,
and the corresponding natural projection
$\Tan Q\times\real\to[\Tan Q\times\real]$,
we prove that both, $\Omega_L^V$ and $E_L^\nabla$,
project to $\omega\in\df^2([\Tan Q\times\real])$ and
 ${\cal E}\in\Cinfty([\Tan Q\times\real])$, in such a way that
$([\Tan Q\times\real],\omega,{\cal E})$ is a regular (symplectic)
Hamiltonian system, although it is not a Lagrangian system, in general,
in spite of $[\Tan Q\times\real]$ being canonically difeomorphic
to a tangent bundle.

The generalization of these results to other more general situations
is a matter of future research. In particular to the following cases:
time-dependent non regular Lagrangian systems,
regular and non-regular time-dependent Lagrangian systems
whose configuration space is a non-trivial fiber bundle
$E\to\real$ (and hence the phase space is a non-trivial jet bundle
 $J^1E\to E\to\real$), and
regular and non-regular classical field theories.


\subsection*{Acknowledgments}

We are grateful for the financial support of the CICYT TAP97-0969-C03-01 and CICTY PB98-0821. We wish to thank Mr. Jeff Palmer for his assistance in preparing the English version of the manuscript.


\end{document}